\documentclass[usenatbib]{mnras}
\usepackage{natbibmnfix, graphicx, times, amsmath, epsfig, amssymb}
\usepackage{verbatim}
\usepackage{color}
\usepackage{mathptmx}
\usepackage{txfonts}
\usepackage[T1]{fontenc}
\usepackage{aecompl}

\def\simgt{\lower.5ex\hbox{\gtsima}} 
\def\simlt{\lower.5ex\hbox{\ltsima}} 
\def\gtsima{$\; \buildrel > \over \sim \;$} 
\def\ltsima{$\; \buildrel < \over \sim \;$}
\def\Msun{M_\odot}

\newcommand\lsim{\mathrel{\rlap{\lower4pt\hbox{\hskip1pt$\sim$}}
        \raise1pt\hbox{$<$}}}
\newcommand\gsim{\mathrel{\rlap{\lower4pt\hbox{\hskip1pt$\sim$}}
        \raise1pt\hbox{$>$}}}
\def\myputfigure#1#2#3#4#5%
{\vskip#5pt\makebox[0pt]{\hskip#2in
\includegraphics[width=#3\textwidth]{#1}}\vskip#4pt\hfill}

\newcommand\lineI[1]{\hbox{#1I}}
\newcommand\lineII[1]{\hbox{#1II}}
\newcommand\lineIII[1]{\hbox{#1III}}

\def\halpha{{\rm H}\alpha}
\def\hbeta{{\rm H}\beta}

\defcitealias{Bowler_2016}{B16}
\newcommand{\zsun}{{\rm Z}_{\odot}}

\title[The nature of the Lyman Alpha Emitter CR7]
      {The nature of the Lyman Alpha Emitter CR7: a persisting puzzle \vspace{-1.5ex}}             
\author[F. Pacucci et al.]
{Fabio Pacucci$^{1,2}$\thanks{fabio.pacucci@yale.edu},
Andrea Pallottini$^{2,3,4}$, Andrea Ferrara$^{2}$, Simona Gallerani$^{2}$\\
$^1$Department of Physics, Yale University, New Haven, CT 06511, USA \\
$^2$Scuola Normale Superiore, Piazza dei Cavalieri, 7  56126 Pisa, Italy \\
$^{3}$Cavendish Laboratory, University of Cambridge, 19 J. J. Thomson Ave., Cambridge CB3 0HE, UK\\
$^{4}$Kavli Institute for Cosmology, University of Cambridge, Madingley Road, Cambridge CB3 0HA, UK\\
\vspace{-10.0ex}}    

\date{}
\pubyear{2017}

\begin{document}
\label{firstpage}
\pagerange{\pageref{firstpage}--\pageref{lastpage}}
\maketitle
    
\begin{abstract}
The peculiar emission properties of the $z \sim 6.6$ Ly$\alpha$ emitter CR7 have been initially interpreted with the presence of either a direct collapse black hole (DCBH) or a substantial mass of Pop III stars. Instead, updated photometric observations by Bowler et al. (2016) seem to suggest that CR7 is a more standard system. Here we confirm that the original DCBH hypothesis is consistent also with the new data.
Using radiation-hydrodynamic simulations, we reproduce the new IR photometry with two models involving a Compton-thick DCBH of mass $\approx 7 \times 10^6 \, \mathrm{\Msun}$ accreting (a) metal-free ($Z=0$) gas with column density $N_H = 8 \times 10^{25} \, \mathrm{cm^{-2}}$, or (b) low-metallicity gas ($Z = 5 \times 10^{-3} \zsun$) with $N_H = 3 \times 10^{24} \, \mathrm{cm^{-2}}$. The best fit model reproduces the photometric data to within $1 \sigma$.
Such metals can be produced by weak star-forming activity occurring after the formation of the DCBH. The main contribution to the \textit{Spitzer}/IRAC $3.6 \, \mathrm{\mu m}$ photometric band in both models is due to \lineI{He}/\lineII{He} $\lambda 4714, 4687$ emission lines, while the contribution of [\lineIII{O}] $\lambda 4959, 5007$ emission lines, if present, is sub-dominant.
Spectroscopic observations with \textit{JWST} will be required to ultimately clarify the nature of CR7.
\end{abstract}


\setcounter{footnote}{1}
\newcounter{dummy}
\section{Introduction}\label{sec:introduction}
Several observations of high-$z$ AGNs (e.g. \citealt{Mortlock_2011} and \citealt{Wu_2015}) have revealed the presence of supermassive black holes (SMBHs, with mass in excess of $10^{9-10} \, \mathrm{\Msun}$) close to the end of the epoch of reionization ($z \sim 7$).
To date, there is no observational evidence of the progenitors of these cosmic behemoths.
Moreover, the detection of $z>6$ SMBHs is in tension with the standard theory of black hole growth.
If Eddington-limited accretion is assumed, the time required to grow such SMBHs from stellar-mass seeds ($\sim 10^2 \,  \mathrm{\Msun}$) born out of the first population of stars (Pop III, \citealt{Haiman_2013}) is longer than the Hubble time at $z = 7$ ($\sim 800 \, \mathrm{Myr}$).

To bypass this time crunch, a possibility is to form SMBHs starting from more massive seeds \citep[$\sim 10^{4-6} \, \mathrm{\Msun}$,][]{Lodato_Natarajan_2006, Devecchi_2009, Volonteri_2010, Davies_2011, Petri12}.
These massive seeds, presumably formed at $z \gsim 10$ \citep{Yue_2014}, can be generated by several mechanisms. In particular, the direct collapse black hole (DCBH) scenario \citep{Bromm_Loeb_2003,Shang_2010, Johnson_2012} is becoming very popular.
The collapse of a metal-free atomic-cooling halo (with a virial temperature $T_{\mathrm{vir}} \gtrsim 10^4 \, \mathrm{K}$) may lead to the formation of DCBHs with a typical mass around $10^5 \, \mathrm{\Msun}$, if the collapse happens in the presence of a strong \citep{Sugimura_2014} flux of Lyman-Werner photons (energy $h\nu =11.2-13.6 \, \mathrm{eV}$) dissociating the $\mathrm{H}_2$ and thus preventing gas fragmentation.

In addition to provide a straight solution to the SMBH growth problem, the DCBH scenario is appealing for several reasons: (i) the model naturally adapts to the physical conditions of the early Universe, (ii) the predicted DCBH mass range may allow detections by current or near-future surveys, and (iii) some galaxy properties (e.g. photometry, gas column density and metallicity, emission line properties) can be used to unveil the presence of a DCBH.
Although no detection of early SMBH progenitors has been confirmed, \cite{Pacucci_2016_DCBH} have proposed that two $z \gsim 6$ objects in the CANDELS/GOODS-S field, also observed in X-rays by Chandra, are the best DCBH candidates to date. 

Interestingly, \cite{Pallottini_Pacucci_2015_CR7} have suggested that a $z \approx 6.6$ object named COSMOS redshift 7 (CR7), the brightest Ly$\alpha$ emitter discovered so far \citep{Sobral_2015,Matthee_2015}, could be powered by a DCBH with an initial mass $\sim 10^5 \, \mathrm{M_{\odot}}$.
In fact, the expected DCBH Spectral Energy Distribution (SED) nicely reproduces the peculiar spectrum of CR7 in its component A (see \citealt{Sobral_2015}). In addition, the DCBH model accounts for the (i) very strong Ly$\alpha$ ($\gsim 8.5 \times 10^{43} \, \mathrm{erg \, s^{-1}}$) and \lineII{He} $\lambda 1640$ ($\sim 2.0 \times 10^{43} \, \mathrm{erg \, s^{-1}}$) emission lines, and the (ii) non-detection (within the instrumental sensitivity) of metal lines. Finally, such a model is also consistent with the X-ray upper limit set by \cite{Elvis_2009} on this source.  Following \cite{Pallottini_Pacucci_2015_CR7}, several works have investigated the possible identification of CR7 with a DCBH \citep{Agarwal_2016,Hartwig_2015, Visbal_2016, Smith_2016, Smidt_2016, Dijkstra_2016_CR7}.
However, very recently \cite{Bowler_2016}, hereafter \citetalias{Bowler_2016}, have obtained deeper observations of CR7 in the optical, near-IR (UltraVISTA DR3) and mid-IR (SPLASH survey) bands, also providing the photometry of the three individual CR7 components A, B and C measured in 3$^{\prime\prime}$ diameter circular apertures on the HST/WFC3 data. The component A of CR7 is the one thought to host the object responsible for the peculiar emission (Pop III stars or DCBH). The other two components, B and C, are instead populated by older (i.e. redder) stars, which might have triggered the formation of the DCBH in the component A by providing the necessary Lyman-Werner flux.
\citetalias{Bowler_2016} claim that the new photometry cannot be reproduced by either a Pop III stellar population or a DCBH synthetic spectra, for which they adopt the \cite{Agarwal_2016} model. Moreover, they suggest that the IRAC $3.6 \, \mathrm{\mu m}$ band might be contaminated by the [\lineIII{O}] $\lambda  4959, 5007$ emission lines. The presence of the [\lineIII{O}] line has been \textit{inferred from the photometry}, and not detected. If confirmed, the presence of metals might be in contrast with the vanilla DCBH scenario described above. As alternative explanations, they propose that CR7 can be classified as a more standard low-mass, narrow-line AGN (which would also explain the lack of radio and X-ray emission) or a young, low-metallicity ($\sim 2 \times 10^{-3} \, \zsun$) starburst with the presence of binaries.

In this \textit{Letter}, we show that the updated photometry of CR7 is well explained by the DCBH model presented by \cite{Pacucci_2015}. Moreover, we provide constraints on the physical properties of this object, as the DCBH mass, accreting gas column density and metallicity. We also discuss how a DCBH in the latest evolutionary phases is essentially indistinguishable from a faint AGN. 

\vspace{-0.6cm}
\section{SED Simulations and Photometry}\label{sec:SED}
We adopt the radiation-hydrodynamic code described in \cite{Pacucci_2015} and \cite{PFVD_2015} to compute the SED of a DCBH. In this Section, we summarize the physical and numerical implementation of our simulations, along with the calculation of the SED and the related photometry. 

\vspace{-0.4cm}
\subsection{Physical framework}
A high-$z$ DCBH, with initial mass $M_{\bullet}(t=0)$, is placed at the centre of a dark matter halo with total mass (baryonic and dark matter) $M_h$ and virial temperature $T_{\mathrm{vir}} = T_{\mathrm{vir}}(M_h,z)  \sim 10^4 \, \mathrm{K}$ \citep{BL01}.
In particular, at $z \sim 10$, $T_{\mathrm{vir}} \sim 10^4 \, \mathrm{K}$ corresponds to $M_h \sim 10^8 \, \mathrm{\Msun}$. The gas component of the host halo initially follows an isothermal density profile $\rho(r) \propto (r/a)^{-2}$, where $a \sim 2 \, \mathrm{pc}$ is the core radius of the baryonic matter distribution. 

Our one-dimensional radiation-hydrodynamic code evolves self-consistently the standard system of ideal, non-relativistic Euler's equations for a gas that is accreting radially onto the central black hole. The initial conditions for the radial velocity are set to a very small, spatially constant value. After a brief transient, much shorter than the free-fall time, the system adjusts to a velocity profile consistent with a rapid accretion across the inner boundary of the grid. The spatial resolution for our simulations is $\sim 10^{-3} \, \mathrm{pc}$.

The black hole accretes mass from the inner regions of the host halo within the transition radius (see \citealt{PVF_2015}): the accretion rate, $\dot{M}_{\bullet}=4 \pi r^2 \rho |v|$ ($v$ is the velocity of the gas) is self-regulated by the combined effects of gravity, gas pressure and radiation pressure.
The accretion rate generates an emitted bolometric luminosity $L_{\rm bol} \equiv \epsilon c^2 \dot{M_{\bullet}}$, where $\epsilon$ is the radiative efficiency of the inflow (varying between $\sim 0.01$ and $\sim 0.1$). The radiation pressure accelerates the gas via $a_{rad}(r) = \kappa(\rho, T) L_{\rm bol}(r)/(4 \pi r^2 c)$, where $\kappa(\rho, T)$ is the gas opacity \citep{Begelman_2008}.
While the standard radiation pressure recipe has been used in our simulations, additional types of radiative feedback could affect, as a second-order term, the environments hosting DCBHs. For instance, \cite{Smith_2017} claim that Lyman-$\alpha$ radiation pressure has a dynamical impact and the presence of radiation-driven winds may also affect the properties of the observed emission line. Moreover, \cite{Inayoshi_2015, Latif_2015, Yue_Pacucci_2016} study the effect of the emitted UV and X-ray photons on the critical flux of Lyman-Werner radiation to form DCBHs. 

While the spherical symmetry is an idealization of a real accretion flow, several works (e.g. \citealt{Novak_2011}) have shown that 1D simulations provide a reliable description of many of its most important features, in particular the accretion rate and duty cycle. The main difference is that the multi-dimensional approach allows to follow the development of fluid instabilities (e.g. Rayleigh-Taylor and Kelvin-Helmholtz). Their net effect is to produce a somewhat less effective feedback and a more irregular pattern of bursts, compared with the 1D case.
The spectral and photometric parts of the calculation, instead, are not expected to be different in a multi-dimensional simulation environment.

To summarize, our code computes the frequency-integrated radiative transfer via a two-stream approximation method coupled with hydrodynamics, also solving the energy equation with appropriate cooling and heating terms. 
We perform two sets of simulations, in which the halo gas has either (i) zero-metallicity ($Z=0$, i.e. primordial composition with a standard helium fraction $Y_P \approx 0.247$), or (ii) $Z= 10^{-4} -  5\times10^{-2} \zsun$, where $\zsun$ is the solar metallicity. See Sec. \ref{sec:fitting_2} for the rationale of a DCBH model extended to a low-metallicity environment. 

\vspace{-0.4cm}
\subsection{SED and photometry}
We post-process the simulation outputs with the code \texttt{CLOUDY} \citep{Cloudy} to compute the frequency-dependent radiative transfer through the host halo, necessary to produce the emerging spectrum. The input data for the \texttt{CLOUDY} code are the following: (i) the spatial profiles for hydrogen number density and temperature, (ii) the source spectrum of the central object, (iii) the bolometric luminosity of the source, and (iv) the gas metal content.
The source spectrum, extending from the far-infrared to hard X-rays, is a standard AGN spectrum \citep{Yue_2014} dependent on the black hole mass, and therefore evolving with time as $M_{\bullet}$ increases.

To compute the photometry of our DCBH model in the required bands, we convolve our synthetic emerging spectrum with the trasmissivity of four wide-band filters used for the component A of CR7 by \citetalias{Bowler_2016}: $Y J_{110}$, $H_{160}$, IRAC-1$(3.6 \, {\rm \mu m})$, IRAC-2$(4.5 \, {\rm \mu m})$.

\vspace{-0.6cm}
\section{Fitting the Photometry of CR7}
In this Section, we compare the DCBH photometry resulting from our model with the one measured by \citetalias{Bowler_2016} for CR7 (component A): the likelihood between the model and the observation is expressed via a $\chi ^2$ probability.
The parameter space has been investigated in the range $10^4 \le M_{\bullet} (\mathrm{M_{\odot}}) \le 5 \times 10^7$, $10^{-4} \le Z (\mathrm{Z_{\odot}}) \le 5\times 10^{-2}$, $10^{24} \le N_H (\mathrm{cm^{-2}})\le 5 \times 10^{26}$ by using 25 logarithmic bins per parameter. The interpolation of the resulting grid is shown in Fig. \ref{fig:M_NH_map_CR7_Z0} and Fig. \ref{fig:M_Z_map_CR7}.

\vspace{-0.4cm}
\subsection{The Zero-Metallicity Case}\label{sec:fitting_1}
\begin{figure*}
\centering
\includegraphics[angle=0,width=0.49\textwidth]{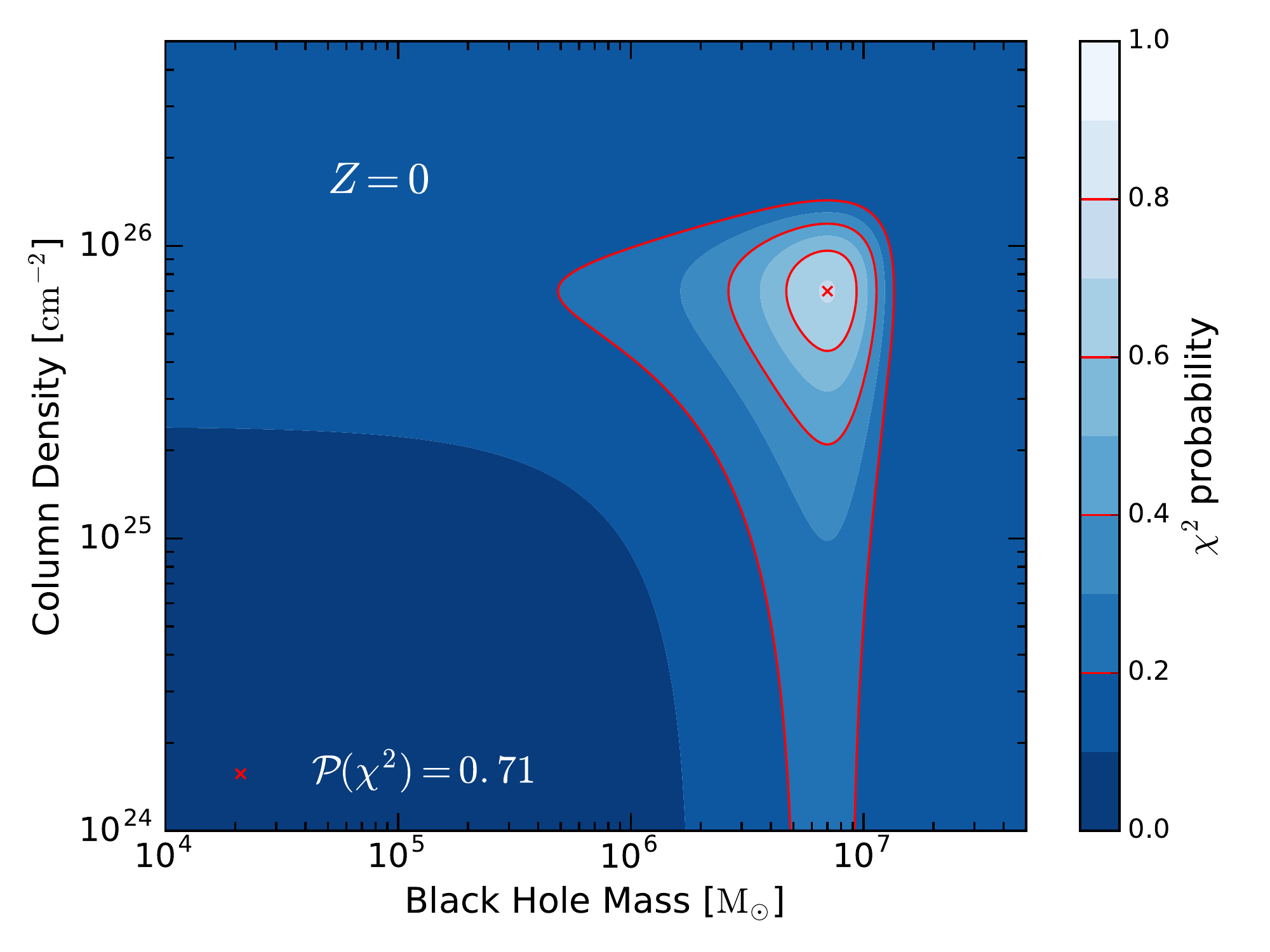}
\includegraphics[angle=0,width=0.49\textwidth]{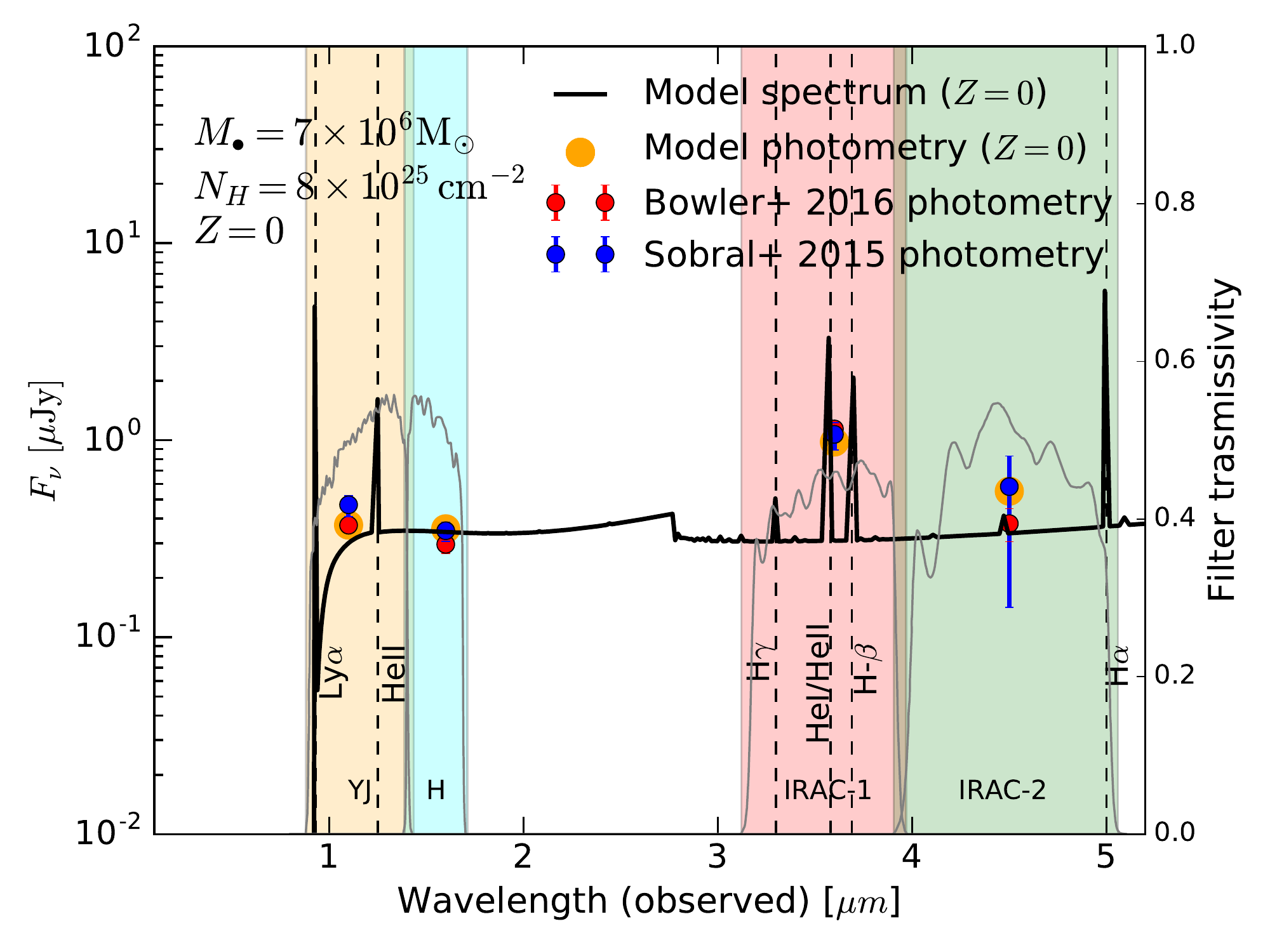}
\vspace{-0.4cm}
\caption{
Zero-metallicity DCBH model.
\textbf{Left panel}: Likelihood between the observed photometry of CR7 (component A) and DCBH model predictions, in the parameter space $M_{\bullet}$-$N_H$. The color scale indicates the $\chi ^2$ probability. The red cross shows our best fit model, reported in the right panel. All models are computed for a metal-free ($Z=0$) gas.
\textbf{Right panel}: Photometry of the best fit (${\cal P}(\chi^2)=0.71$) $Z=0$ model. The best fit is obtained with a black hole mass $\sim 7 \times 10^6 \, \mathrm{\Msun}$ and a column density $N_H \sim 8 \times 10^{25} \, \mathrm{cm^{-2}}$. All the emission lines are produced by H or He atoms (note the contribution of \lineI{He}/\lineII{He}  $\lambda 4714, 4687$ lines in the IRAC-1 band). The transmissivity of the filters employed is shown.
\label{fig:M_NH_map_CR7_Z0}
}
\end{figure*}

The parametric study in the space $M_{\bullet}$-$N_H$ is shown in the left panel of Fig. \ref{fig:M_NH_map_CR7_Z0}, while the best fit model is shown in the right panel. With this model, we obtain a fairly good best fit, i.e. ${\cal P}(\chi^2)=0.71$.

For $Z=0$, the main driver of the IRAC-1 infrared excess is the \lineI{He} line at $\lambda = 4714 \, \AA$ rest-frame, which falls roughly at the centre of the $3.6 \, \mathrm{\mu m}$ band for an object at $z \sim 6.6$, followed by the H$\beta$ line. Such He line corresponds to the transition from  level $1s2p$ to level $1s4s$. There might be also a contribution from the \lineII{He} line at $\lambda = 4687\, \AA$ rest-frame, which also falls very close to the \lineI{He}~$\lambda 4714$ line.
In the IRAC-2 band the main contribution comes from the strong H$\alpha$ line.
This result clearly shows that the contribution from the [\lineIII{O}] line is not necessary, in contrast with the statement made by \citetalias{Bowler_2016} that the new photometry of CR7 cannot be fitted by current DCBH models, for which they adopted the \cite{Agarwal_2016} model. 
Our DCBH model, instead, is able to reproduce the new photometry with a good accuracy, in a $Z=0$ environment, in overall agreement with the classical model for DCBH formation.
The best fit is obtained with a black hole of $\sim 7 \times 10^6 \, \mathrm{\Msun}$, reached in $\sim 97 \, \mathrm{Myr}$ of accretion onto a DCBH of initial mass $\sim 10^5 \, \mathrm{\Msun}$.

In order to fit the new photometry of CR7 our model requires a large absorbing gas column density, $N_H \sim 8 \times 10^{25} \, \mathrm{cm^{-2}}$, implying that CR7 should be Compton-thick. This explains why CR7 remains undetected in deep Chandra observations \citep{Elvis_2009}. The large column density requirement is set by the high measured IRAC-1 infrared flux (powered by the two-photon emission process) resulting in a blue color, [3.6]-[4.5] = -0.74. This value is consistent within $\simeq1.5\sigma$ with the color measured by \citetalias{Bowler_2016}, i.e. $-1.2\pm 0.3$. The $N_H$ value is in agreement with predictions based on limits set by the unresolved fraction of the X-ray background \citep{Yue_2014}. Such population of highly-obscured ($N_{\rm H}\simgt 10^{25} \rm cm^{-2}$) DCBHs has been invoked to explain the observed level of the infrared background fluctuations.
To conclude, the existing CR7 data is reproducible with fair accuracy by a vanilla Compton-thick DCBH model in which the accreting gas is primordial.

\vspace{-0.4cm}

\subsection{The Low-Metallicity Case}
\label{sec:fitting_2}
The presence of a pristine or nearly-pristine gas is a pre-requisite for the \textit{formation} of a DCBH. Nonetheless, the DCBH host halo might eventually merge with other metal-polluted halos, or even witness some weak star formation episodes in the accretion flow onto the newly formed DCBH. Hence, the presence of a small amount of metals in a halo does not necessarily imply that a DCBH cannot be present. 
Since the cosmic era during which the formation of DCBHs is more likely to occur is in the redshift range $13 \lsim z \lsim 20$ \citep{Yue_2014}, at much later epochs it is conceivable that halos originally harboring DCBHs have become polluted. \cite{Natarajan_2016} already pointed out that the merger of a pristine halo harboring a newly-formed DCBH with a metal-polluted halo would generate such an intermediate object.
As a matter of fact, the number of metal-free halos decreases rapidly between $z\sim 6$ and $z\sim 4$ \citep[e.g.][]{Pallottini_2014}. This could be the case of CR7 that, since the era when its central DCBH possibly formed, by $z \sim 6.6$ could have experienced mergers and/or accrete material polluted by external galaxies, that typically rise the metallicity of the intergalactic medium to $Z\sim 10^{-3}\zsun$ \citep[][]{Pallottini_2014, Smith_metal_2015}.

\cite{Natarajan_2016} investigated how the presence of a small amount of metals ($10^{-3} \, \zsun$-$10^{-2} \, \zsun$) modifies the spectrum of the source. Such analysis shows that variations of metallicity and column density are degenerate: it is possible to obtain the same optical depth to high-energy photons with a small column ($N_H \lesssim 10^{25} \, \mathrm{cm^{-2}}$) of enriched gas ($Z \gsim 0.01 Z_{\odot}$), or with a larger column ($N_H \gsim 10^{25} \, \mathrm{cm^{-2}}$) of lower metallicity gas ($Z \lsim 0.01 Z_{\odot}$).
The high-energy ($E \gsim 1 \, \mathrm{keV}$) absorption translates into a higher infrared re-emission, due to different physical processes: the two-photon emission is increased by a larger absorbing column density, while Auger-like processes are enhanced by a higher metallicity.
Because of such degeneration, it is possible to fit the photometry of CR7 with a model that has a column density lower than the ones investigated in Sec. \ref{sec:fitting_1} and a non-negligible metallicity.

\begin{figure*}
\centering
\includegraphics[angle=0,width=0.49\textwidth]{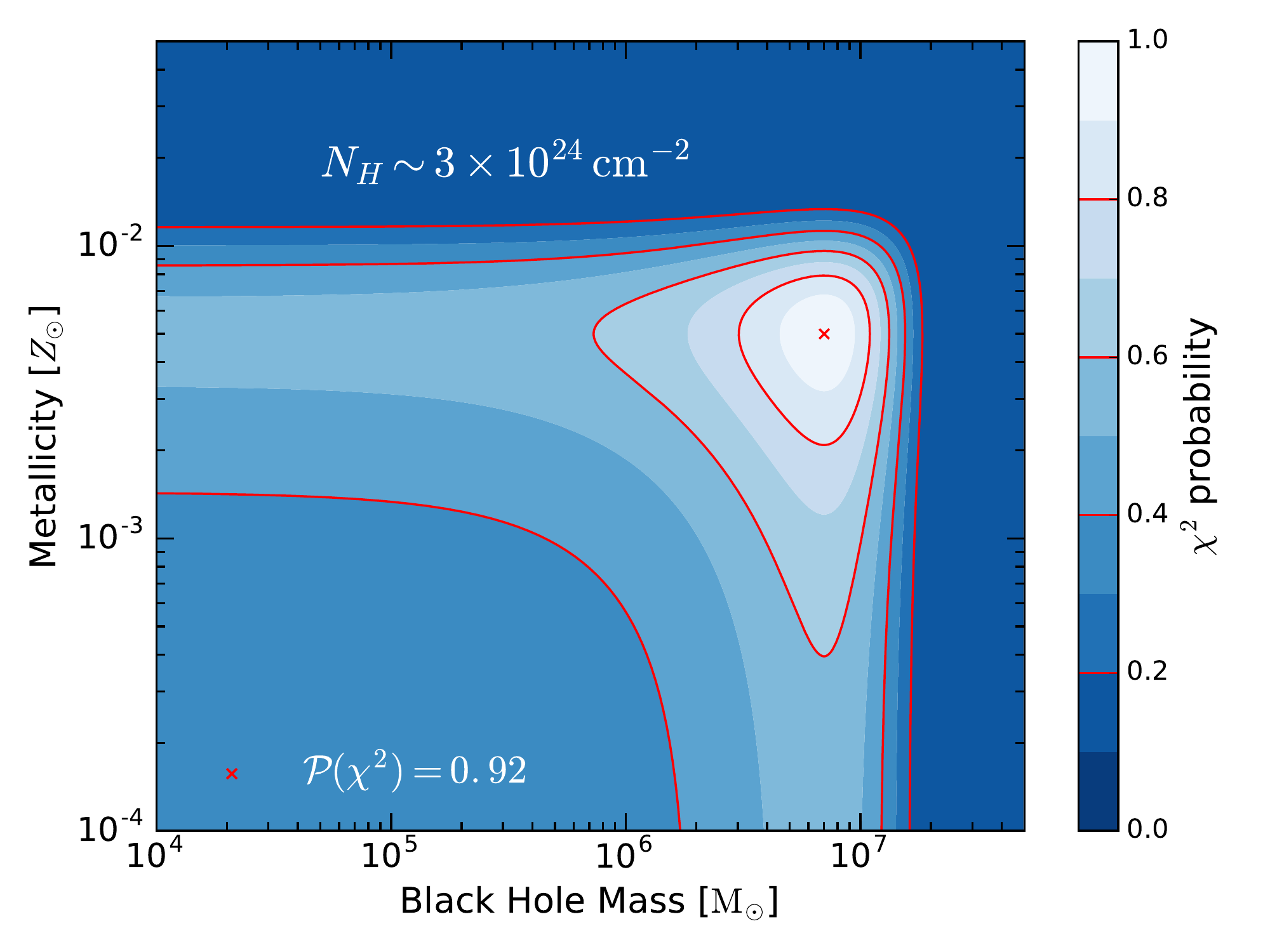}
\includegraphics[angle=0,width=0.49\textwidth]{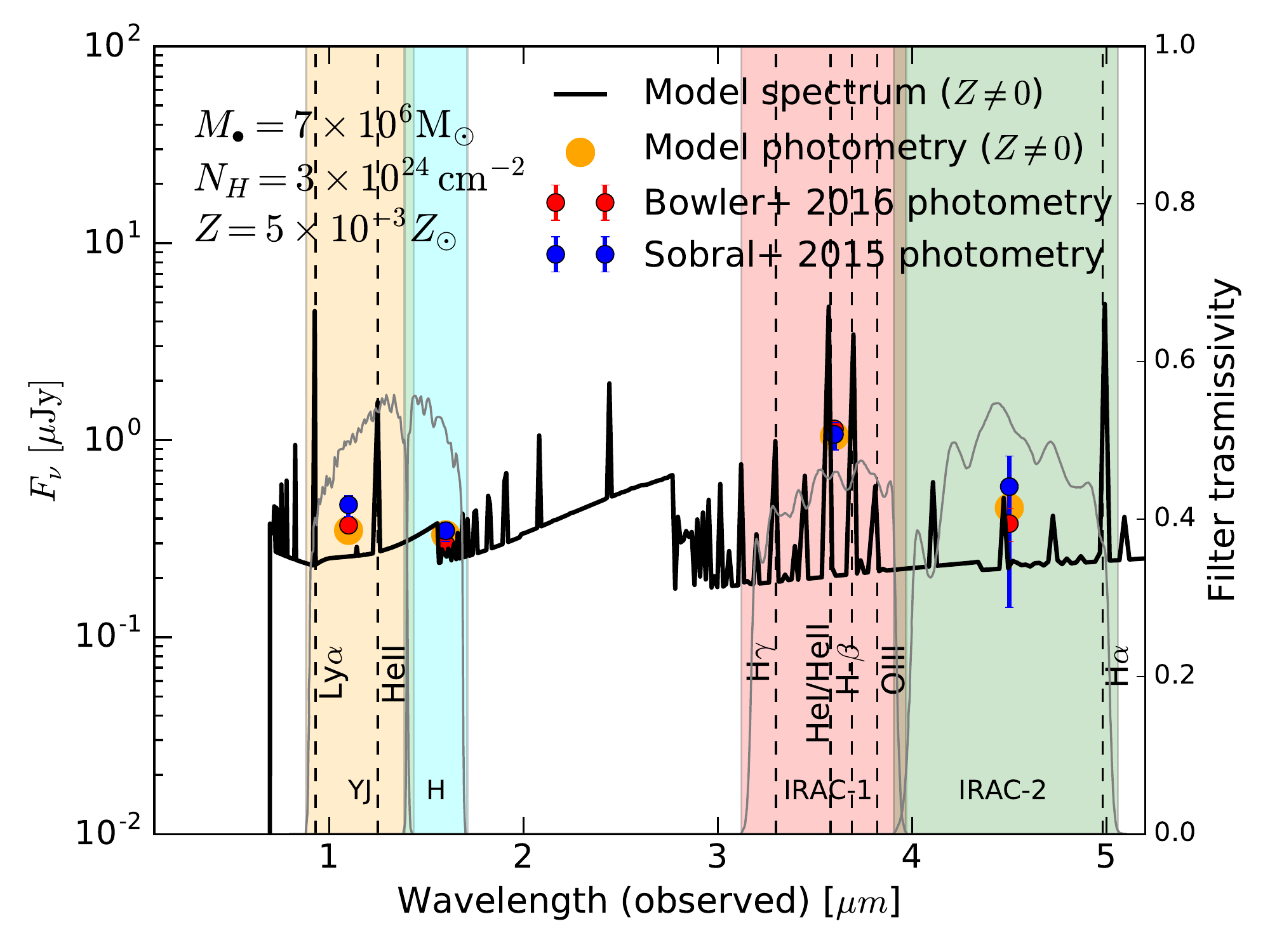}
\vspace{-0.4cm}
\caption{
As in Fig. \ref{fig:M_NH_map_CR7_Z0}, for the low-metallicity DCBH model. 
\textbf{Left panel}: Likelihood between the photometry of CR7 (component A) and the prediction for the DCBH model, in the parameter space $M_{\bullet}-Z$.
The red cross shows our best fit model (see right panel). All these models are computed for a fixed gas column density $N_H \sim 3 \times 10^{24} \, \mathrm{cm^{-2}}$.
\textbf{Right panel:} SED of the best fit model for the low-metallicity parameter space shown in the left panel. The model has a likelihood ${\cal P}(\chi ^2) = 0.92$ and corresponds to $M_{\bullet} \sim 7 \times 10^6 \, \mathrm{\Msun}$ and $Z \sim 5 \times 10^{-3} \zsun$. The emission lines more energetic than the Ly$\alpha$, visible here, are absent in the $Z=0$ model because in that case the value of $N_H$ is much larger and these emission lines become optically thick.
\label{fig:M_Z_map_CR7}
}
\end{figure*}
The photometry dependence on $M_{\bullet}$ and $Z$ is shown in the left panel of Fig. \ref{fig:M_Z_map_CR7}. The best fit model (Fig. \ref{fig:M_Z_map_CR7}, right panel) is characterized by $M_{\bullet} \sim 7 \times 10^6 \, \mathrm{\Msun}$, $Z \sim 5 \times 10^{-3} \zsun$ and ${\cal P}(\chi ^2) = 0.92$. The mass is reached after $\sim 92 \, \mathrm{Myr}$ of accretion onto a DCBH of initial mass $\sim 10^5 \, \mathrm{\Msun}$.
The gas column density is assumed to be $N_H \sim 3 \times 10^{24} \, \mathrm{cm^{-2}}$, in line with theoretical predictions for DCBHs (see e.g. the discussion in \citealt{PVF_2015}) and in accordance with our previous estimate for CR7 \citep{Pallottini_Pacucci_2015_CR7}.
Such small metal pollution makes it possible to reproduce the observed infrared emission in the $3.6 \, \mathrm{\mu m}$ and $4.5 \, \mathrm{\mu m}$ bands and it is compatible with the upper limits set by \cite{Sobral_2015}: \lineII{He}/[\lineIII{O}]$\lambda 1663$ $ > 3$ and \lineII{He}/\lineIII{C}]$\lambda 1908$ $ >2.5$. 
Noticeably, the \lineI{He}/\lineII{He}  $\lambda 4714, 4687$ lines fall at the centre of the IRAC-1 band for an object at $z \sim 6.6$: their contribution to the photometry at $3.6 \, \mathrm{\mu m}$ is then maximal. The low-metallicity gas in our model does not result in very strong metal lines. 
While in our models the [\lineIII{O}] line is severely sub-dominant with respect to the \lineI{He}/\lineII{He} lines, in \citetalias{Bowler_2016} such metal line (along with H$\beta$) is advocated to explain the excess in the $3.6 \, \mathrm{\mu m}$ band.
%
%
Given the canonical $\halpha/\hbeta \sim 2.87$ ratio \citep{Osterbrock_2006}, the expected value of the $[3.6]-[4.5]$ colour would be close to zero, if no additional contributions from nebular lines were present. The blue ($[3.6]-[4.5] \sim -1.2$) colour measured by \citetalias{Bowler_2016} implies the presence of a nebular line in the IRAC-1 band. This has been identified by \citetalias{Bowler_2016} as the [\lineIII{O}] line yielding a total $EW_0(\hbeta + [\lineIII{O}]) \gsim 2000 \, \AA$, also following previous inferences (e.g. \citealt{Osterbrock_2006}).  We have shown instead that \lineI{He}/\lineII{He} lines arising in a metal-free or low-$Z$ gas can equally well explain the observed photometry. 
Because of the absence of high-resolution spectra in the IRAC-1 and IRAC-2 bands, it is currently impossible to discriminate between the dominance of metal lines or \lineI{He}/\lineII{He} lines.
A spectroscopic analysis of this source carried out with \textit{JWST} in these bands will be necessary to understand the nature of the energy source of CR7. 

Finally, we predict a \lineII{He} $\lambda 1640$/Ly$\alpha \sim 0.31$ line ratio and an \lineII{He} $\lambda 1640$ equivalent width of $\sim 88 \, \AA$. Such values well match the ones measured by \cite{Sobral_2015}, i.e. a ratio of $0.23\pm 0.1$ and $EW_0(\lineII{He}\lambda 1640)=80 \pm 20 \, \AA$. The $EW_0(\lineII{He}\lambda 1640)$ predicted by our model is larger than the one reported by \citetalias{Bowler_2016}, but still consistent within $1.6\sigma$.
To conclude, our model can reproduce with excellent accuracy the infrared photometry of CR7 with a DCBH accreting gas with a largely sub-solar ($\sim 5 \times 10^{-3} \zsun$) metallicity, and hence, without strong metal line contamination of the IRAC bands.

\vspace{-0.6cm}
\section{DCBH and faint AGN at $z \lesssim 6$: just semantics?}\label{sec:DCBH_AGN}
In \citetalias{Bowler_2016} the possibility that CR7 is powered by a DCBH was discarded in favor of a more classical low-mass, narrow-line AGN model or, in other words, a faint AGN.

Faint AGNs are objects of great interest, since they may play a major role in the process of reionization \citep{Giallongo_2015} and they could also have important implications concerning the abundance and mass of SMBH seeds and their early growth \citep{Volonteri_2012}. 
In the most common classification of faint AGNs, they are point-like sources detected in the X-rays and with an overall X-ray luminosity ($2-10 \, \mathrm{keV}$) $L_X \lsim 10^{44} \, \mathrm{erg \, s^{-1}}$ \citep{Giallongo_2015}. The black holes powering these sources are either intermediate-mass objects ($\sim 10^{5} \, \mathrm{\Msun}$) accreting at nearly the Eddington rate, or SMBHs ($\gsim 10^{7} \, \mathrm{\Msun}$) accreting at very low rates ($\dot{M}/\dot{M}_{\rm Edd} \lsim 0.01$).
The same geometry of the accretion disk surrounding these objects largely varies depending on the accretion rate. For objects accreting at nearly the Eddington rate, a standard $\alpha$-disk (\citealt{Shakura_Sunyaev_1976}) model should be realistic. For objects with $\dot{M}/\dot{M}_{\rm Edd} \lsim 0.01$, the accretion disk should inflate and become advection dominated (or ADAF, advection dominate accretion flows, \citealt{Rees_1982}). Also the absorbing column densities of these objects may largely vary as a nearly independent parameter, affecting their X-ray output.
%
Thus, the category of faint AGN encompasses a large group of objects, with very different features and with the only common observational constraint of having a relatively low X-ray luminosity.

The theoretically predicted physical properties of DCBHs are rather different. They are expected to be objects of low or intermediate mass ($<10^{6} \, \mathrm{\Msun}$ at their formation), accreting at or above the Eddington rate and with very large absorbing column densities ($\gsim 10^{25} \, \mathrm{cm^{-2}}$, see e.g. \citealt{PVF_2015} and references therein).

With particular regard to the low-redshift Universe ($z \lesssim 6$), the observational discrimination between DCBHs and faint AGNs is far from being trivial.
An astrophysical black hole is completely defined by its mass (and spin). However, information of its seed mass is eventually washed out by the accretion history: hints on its origin must be extracted from the properties of the host galaxy.
In general, while a DCBH is a newly formed object, a faint AGN is an evolved one.
A DCBH could be considered to enter the AGN category when a substantial stellar component merges with or forms into the host halo, or because external pollution is in place. Conversely, an AGN has to be interpreted as an advanced stage of the black hole evolution, that started from a seed that may have formed by a plethora of different processes.
Analyzing the combined properties of the black hole and its host halo, in any moment between the seed stage and the AGN stage, it could be possible to understand which process drove the formation of the original seed. The main discriminant between a newly formed DCBH and a more evolved (faint) AGN would be the gas metallicity of the host halo.
The time needed for a black hole seed of $\sim 10^5 \, \mathrm{\Msun}$ to reach the predicted mass for CR7 is approximately $\sim 95 \, \mathrm{Myr}$ in both our models. 
Star formation activity should have started on a much shorter time scale, about $5 \, \mathrm{Myr}$ for a typical molecular cloud of density $\sim 100 \, \mathrm{cm^{-3}}$. Nonetheless, the current presence of a Lyman-Werner flux (generated by the components B and C) above the critical threshold (see \citealt{Pallottini_Pacucci_2015_CR7}) should have delayed it so far. Hence, major star forming activity is not expected to have occurred in the galaxy hosting CR7. Episodic star formation may have been triggered by external pollution or minor mergers, and could have manifested itself in the small amount of metals predicted in our low-$Z$ model.

To conclude, the real question is not if CR7 is a DCBH or a low-mass, narrow-line AGN, but whether it hosts a black hole and, if so, which process is responsible for the formation of its original seed. 
An alternative to the DCBH interpretation may be a Pop III starburst \citep{Sobral_2015, Visbal_2016}, albeit this scenario encounters many difficulties \citep{Yajima_2016}.

\vspace{-0.6cm}
\section{Conclusions}\label{sec:conclusions}

By performing a series of radiation-hydrodynamic simulations of the accretion process onto high-$z$ black hole seeds, we have showed that the updated photometry of the $z \sim 6.6$ Ly$\alpha$ source CR7 (see \citealt{Sobral_2015} and \citetalias{Bowler_2016}) is well reproduced by a DCBH model. In particular, we are able to reproduce the CR7  photometry in the YJ, H, IRAC-1 and IRAC-2 bands with high fidelity. After this work was completed, \cite{Agarwal_2017} proposed a model similar to our $Z \ne 0$ scenario, based on a $\sim 10^6 \, \mathrm{\Msun}$ black hole in a $0.01 \, \mathrm{\zsun}$ metallicity environment. We note that, while our results are consistent with the \citetalias{Bowler_2016} photometry to within $1\sigma$, \cite{Agarwal_2017} fails to reproduce the IRAC-2 photometry at $\sim 3\sigma$ level.

Since the gas metallicity effect on the infrared photometry is degenerate with its column density, we investigated two models of a Compton-thick DCBH accreting (a) metal-free and (b) low-metallicity gas.
For $Z=0$, we found a best fit model for which $N_H \sim 8 \times 10^{25} \, \mathrm{cm^{-2}}$. In the low-metallicity environment the best solution implies $Z \sim 5 \times 10^{-3} \zsun$, assuming $N_H \sim 3 \times 10^{24} \, \mathrm{cm^{-2}}$. In both cases, the DCBH mass is $\sim 7 \times 10^{6} \, \mathrm{\Msun}$, reached after $\sim 95 \, \mathrm{Myr}$ of accretion, in overall agreement with other previous predictions, e.g. \cite{Dijkstra_2016_CR7}. This value is also consistent with the maximum mass achievable by a black hole accreting in an isolated halo, predicted in \cite{Pacucci_2017_maxmass}. While the black hole mass affects the luminosity of the source, the overall photon distribution of the outgoing spectrum is mainly controlled by the metallicity and the column density.
The net photometry of the accreting object is degenerate in the two-dimensional parameter space of metallicity and column density, but in principle this degeneracy might be broken by the addition of some spectral information, like the Ly$\alpha$ and the \lineII{He} line properties. In fact, their overall output should be mainly affected by the column density and not by the metallicity of the gas. Since the goal of this Letter is to show that on its own the current photometry of CR7 is insufficient to discriminate between different scenarios, we postpone this investigation to future work.

While \citetalias{Bowler_2016} suggested that the main contribution to the $3.6 \, \mathrm{\mu m}$ photometry should come from the [\lineIII{O}] emission line, our model shows instead that \lineI{He}/\lineII{He} lines are the main contributors in both cases, i.e. the [\lineIII{O}] emission, if present, is sub-dominant. 
The strong helium line has been associated with both a Pop III starburst \citep{Sobral_2015} or with the DCBH scenario \citep{Pallottini_Pacucci_2015_CR7}. Nonetheless, it is also a characteristic of low-luminosity AGNs in highly star-forming galaxies \citep{HeII_2016} and of the narrow-line region of low-mass AGNs \citep{Ludwig_2012}. 

To conclude, we discussed and compared the different properties of DCBHs and faint AGNs. We suggested that at $z \lesssim 6$ the differences between these classes of objects are blurred, since AGNs are fundamentally a later stage of the evolution of DCBHs. The measurement of the (alleged) black hole mass, absorbing column density and, most importantly, metallicity of the host halo will classify CR7 as a member of one of these two classes of objects.
Deep spectroscopic observations, likely using the NIRSpec instrument aboard \textit{JWST}, are required to finally shed light on the CR7 puzzle. 

\vspace{0.3cm}
We thank Rebecca Bowler and Eli Visbal for their valuable comments. F.P. acknowledges the ADAP grant No. MA160009.

\vspace{-0.5cm}
\bibliographystyle{mnras}
\bibliography{ms}
\label{lastpage}
\end{document}